\documentclass[10pt,twocolumn,american,10pt,twocolumn,aps,superscriptaddress,subscriptfootnote,nofootinbib]{revtex4}
\usepackage[utf8]{inputenc}
\usepackage{amsmath}
\usepackage{amssymb}
\usepackage{graphicx}

\makeatletter
\@ifundefined{textcolor}{}
{%
 \definecolor{BLACK}{gray}{0}
 \definecolor{WHITE}{gray}{1}
 \definecolor{RED}{rgb}{1,0,0}
 \definecolor{GREEN}{rgb}{0,1,0}
 \definecolor{BLUE}{rgb}{0,0,1}
 \definecolor{CYAN}{cmyk}{1,0,0,0}
 \definecolor{MAGENTA}{cmyk}{0,1,0,0}
 \definecolor{YELLOW}{cmyk}{0,0,1,0}
}

\usepackage[english,portuguese]{babel}

\makeatother

\usepackage{babel}
\begin{document}

\title{The Maccone-Pati uncertainty relation}

\author{Jonas Maziero}

\email{jonas.maziero@ufsm.br}

\address{Departamento de Física, Centro de Ciências Naturais e Exatas, Universidade
Federal de Santa Maria, Avenida Roraima 1000, 97105-900, Santa Maria,
RS, Brazil}
\begin{abstract}
The existence of incompatible observables constitutes one of the most
prominent characteristics of quantum mechanics (QM) and can be revealed
and formalized through uncertainty relations. The Heisenberg-Robertson-Schrödinger
uncertainty relation (HRSUR) was proved at the dawn of quantum formalism
and is ever-present in the teaching and research on QM. Notwithstanding,
the HRSUR possess the so called triviality problem. That is to say,
the HRSUR yields no information about the possible incompatibility
between two observables if the system was prepared in a state which
is an eigenvector of one of them. After about 85 years of existence
of the HRSUR, this problem was solved recently by Lorenzo Maccone
and Arun K. Pati. In this article, we start doing a brief discussion
of general aspects of the uncertainty principle in QM and recapitulating
the proof of HRSUR. Afterwards we present in simple terms the proof
of the Maccone-Pati uncertainty relation, which can be obtained basically
via the application of the parallelogram law and Cauchy-Schwarz inequality.
\\
 \textbf{Keywords:} Quantum mechanics, uncertainty relations 
\end{abstract}
\maketitle

\section{Introduction}

One can say that uncertainty is an integral part of our lives \cite{Mlodinow}.
However, the uncertainties we face in our daily lives are frequently
something associated more with our ignorance as observers than a characteristic
property of the physical entities with which we interact. This scenario
changes completely in situations where quantum effects are observationally
important. For these systems uncertainty is a fundamental character.
That is to say, we just cannot, in general, foretell what is going
to happen in the future, even if we have all the information we can
have about the history of the object we are describing \cite{Bell,Aspect,Hensen,Giustina,Shalm}.

For systems whose description requires the use of quantum mechanics
(QM) \cite{Griffiths,Sakurai,Maziero_rbef15,Maziero_rbef16}, we can
only calculate probabilities (chances or relative frequencies) for
an event to occur. This fact can be attributed to the existence, in
QM, of incompatible observables (IO). Once these observables are represented
by non-commuting Hermitian matrices, which as a consequence cannot
share all eigenvectors, we can, via the measurement of one of them,
prepare a state that is a superposition of the eigenvectors of the
other observable. In this case, the uncertainty about this last observable
is necessarily non-null. This is associated with a positive ``width''
(measured using e.g. the standard deviation or variance) of the probability
distribution (PD) for its eigenvalues.

If we prepare a physical system in state $|\xi\rangle$ via the measurement
of an observable $\hat{C}$ \cite{Maziero_rbef15}, we can utilize
the kinematic structure of QM to derive restrictions on how small
can be the product or sum of the uncertainties associated with other
two observables $\hat{A}$ and $\hat{B}$ \cite{Heisenberg,Robertson,Schrodinger,Maccone-Pati}.
This kind of inequality, which is dubbed \textit{preparation uncertainty
relation} (PUR), depends on $|\xi\rangle$ and on the regarded observables
and is the main theme of this article.

The goal of a PUR is to identify (and somehow quantify) the state-dependent
incompatibility of two observables via the general impossibility of
preparing the physical system of interest in a state for which both
probability distributions (for the eigenvalues of these observables)
have null variance. The frequent presence of this kind of uncertainty
relation (UR) in QM textbooks points towards its didactic importance
concerning the learning of the fundamentals of this theory. Moreover,
UR have diverse practical applications, going from the justification
for the use of a complex field in QM \cite{Maczynski} to areas such
as quantum cryptography \cite{Koashi} and entanglement witness \cite{Takeuchi,Guhne}.

There are several other relevant aspects of the uncertainty principle
of QM \cite{Busch-1}, and we shall mention some of them in this paragraph.
In Quantum Information Science \cite{Nielsen}, especially in Quantum
Cryptography \cite{Wilde}, error-disturbance UR are particularly
important because they impose limits on the amount of information
we can obtain by making measurements in a system and the consequent
disturbance which will be impinged on its state \cite{Ozawa,Branciard,Wilde-1}.
It is worth mentioning that, as in measuring an observable to extract
information about the system we shall generally modify the PD of another
observable, the error-disturbance UR are closely related to the UR
for joint measurement of these observables. On the other hand, the
recognition that quantum correlations, such as entanglement \cite{Horodecki,Davidovich}
and discord \cite{Maziero_ijqi,Vedral}, can be utilized as a resource
for the more efficient manipulation of information motivated the proposal
and analysis of UR with quantum memories \cite{Renner,Wilde-2}. Here
it has been shown that the constraints on the variances of IO of a
system can be weakened if the observer is quantumly correlated with
it. Besides, entropic UR, independent of state, can be obtained which
constrain the ``entropies'' of the PD of IO \cite{Wehner}. Another
important kind of UR are those involving parameters which are not
represented by Hermitian operators, such as time or phase \cite{Caves}.
An example of this kind of UR is the energy-time UR, which has a fundamental
role for proving limits on how fast quantum states can change with
time; which by its turn can be utilized to limit the efficiency of
quantum information processing devices \cite{Frey}. It is worthwhile
observing that as the majority of the UR mentioned above involve the
measurement of the average of the product of to IO, $\hat{A}\hat{B}$,
which is not an Hermitian operator, they are not amenable for experimental
tests \cite{Busch}. Recently a scheme has been proposed which can,
in principle, turn possible the experimental verification of UR involving
the average value of $\hat{A}\hat{B}$ \cite{Vedral1,Vedral2}, but
such a technique has not been put to work yet.

The sequence of this article is organized as follows. In Sec. \ref{sec:RIHRS}
we discuss the Cauchy-Schwarz inequality and its use for obtaining
the UR of Heisenberg, Robertson, and Schrödinger (HRSUR). Afterwards
we discuss the triviality problem of the HRSUR and prove, in Sec.
\ref{sec:RIMP}, the UR of Maccone and Pati (MPUR). In contrast to
the HRSUR, the MPUR leads to non-zero lower bounds for the sum of
the variances of two observables whenever the system state is not
an eigenvector of both corresponding Hermitian operators; therefore
the MPUR can be seen as an improvement for the HRSUR. At last, after
presenting an example of application of these uncertainty relations
in Sec. \ref{sec:exemplos}, some final remarks are included in Sec.
\ref{sec:conclusoes}.

\section{Heisenberg-Robertson-Schrödinger uncertainty relation and its triviality
problem}

\label{sec:RIHRS}

In view of its importance for proving the results we discuss in this
article, we shall begin recapitulating the Cauchy-Schwarz inequality
(CSI). The CSI states that for any pair of non-null vectors $|\psi\rangle$
and $|\phi\rangle$ in a Hilbert space $\mathcal{H}$ \cite{Maziero_rbef15},
it follows that 
\begin{equation}
\langle\psi|\psi\rangle\langle\phi|\phi\rangle\ge|\langle\psi|\phi\rangle|^{2},\label{eq:DCS}
\end{equation}
with equality obtained if and only if $|\psi\rangle$ and $|\phi\rangle$
are collinear. Let us recall that, for the state spaces we deal with
here, the inner product between two vectors $|\psi\rangle$ and $|\phi\rangle$
is defined as: $\langle\psi|\phi\rangle=|\psi\rangle^{\dagger}|\phi\rangle$,
with $x^{\dagger}$ being the conjugate transpose of the vector (or
matrix) $x$. We observe that a simple manner to prove the CSI is
by applying the positivity of the norm, 
\begin{equation}
{\normalcolor \parallel}|\xi\rangle{\normalcolor \parallel}=\sqrt{\langle\xi|\xi\rangle}\ge0,
\end{equation}
to the vector $|\xi\rangle=|\psi\rangle-(\langle\phi|\psi\rangle/\langle\phi|\phi\rangle)|\phi\rangle$.
The condition for equality in the CSI can be inferred from the fact
that ${\normalcolor \parallel}|\xi\rangle{\normalcolor \parallel}=0$
if and only if $|\xi\rangle$ is the null vector.

Let us see how the CSI can be used for deriving the Heisenberg-Robertson-Schrödinger
uncertainty relation (HRSUR). Let $\hat{A}$ and $\hat{B}$ be two
observables of a physical system prepared in state $|\xi\rangle$.
Let $\langle\hat{X}\rangle=\langle\xi|\hat{X}|\xi\rangle$ denote
the average value of any operator $\hat{X}$, and we use $\mathbb{I}$
for the identity operator in $\mathcal{H}$. Then we define the vectors
\begin{equation}
|\psi\rangle=(\hat{A}-\langle\hat{A}\rangle\mathbb{I})|\xi\rangle\mbox{ and }|\phi\rangle=(\hat{B}-\langle\hat{B}\rangle\mathbb{I})|\xi\rangle\label{eq:states}
\end{equation}
and substitute them in the CSI. Firstly we notice that 
\begin{equation}
\langle\psi|\psi\rangle=\mathrm{Var}(\hat{A})\mbox{ and }\langle\phi|\phi\rangle=\mathrm{Var}(\hat{B}),
\end{equation}
with $\mathrm{Var}(\hat{X})=\langle(\hat{X}-\langle\hat{X}\rangle\mathbb{I})^{2}\rangle$
being the variance of $\hat{X}$. We can also verify that 
\begin{eqnarray}
\langle\psi|\phi\rangle & = & \langle\hat{A}\hat{B}-\langle\hat{A}\rangle\langle\hat{B}\rangle\mathbb{I}\rangle\\
 & = & 2^{-1}\langle\{\hat{A},\hat{B}\}-2\langle\hat{A}\rangle\langle\hat{B}\rangle\mathbb{I}\rangle+2^{-1}\langle[\hat{A},\hat{B}]\rangle,\nonumber 
\end{eqnarray}
where 
\begin{equation}
[\hat{A},\hat{B}]=\hat{A}\hat{B}-\hat{B}\hat{A}\mbox{ and }\{\hat{A},\hat{B}\}=\hat{A}\hat{B}+\hat{B}\hat{A}
\end{equation}
are, respectively, the commutator and anti-commutator of $\hat{A}$
and $\hat{B}$. As $\{\hat{A},\hat{B}\}$ and $[\hat{A},\hat{B}]$
are, respectively, Hermitian and anti-Hermitian operators, their mean
values are, respectively, purely real and purely imaginary numbers.
So, considering that 
\begin{equation}
|\langle\psi|\phi\rangle|^{2}=(\mathrm{Re}\langle\psi|\phi\rangle)^{2}+(\mathrm{Im}\langle\psi|\phi\rangle)^{2},
\end{equation}
after some manipulations we obtain the HRSUR \cite{Heisenberg,Robertson,Schrodinger}%
\footnote{Even though this inequality is usually dubbed Heisenberg's uncertainty
relation, here we prefer to give credit also for Robertson and Schrödinger,
who have obtained it in its more general forms. An alternative proof
of HRSUR can be found in Ref. \cite{Rigolin}.%
}: 
\begin{equation}
\mathrm{Var}(\hat{A})\mathrm{Var}(\hat{B})\ge(\mathrm{CovQ}(\hat{A},\hat{B}))^{2}+2^{-2}|\langle[\hat{A},\hat{B}]\rangle|^{2}=T_{1},\label{eq:RIHRS}
\end{equation}
where
\begin{equation}
\mathrm{CovQ}(\hat{A},\hat{B})=2^{-1}(\mathrm{Cov}(\hat{A},\hat{B})+\mathrm{Cov}(\hat{B},\hat{A}))
\end{equation}
is the quantum covariance, with $\mathrm{Cov}(\hat{X},\hat{Y})=\langle\hat{X}\hat{Y}\rangle-\langle\hat{X}\rangle\langle\hat{Y}\rangle$
being the covariance between the observables $\hat{X}$ and $\hat{Y}$.
If $\hat{A}$ and $\hat{B}$ are compatible observables, i.e., if
$[\hat{A},\hat{B}]=\hat{0}$, then we shall have $\mathrm{CovQ}(\hat{A},\hat{B})=\mathrm{Cov}(\hat{A},\hat{B})$.

Let us look now at the \emph{triviality problem} of the HRSUR. Without
loss of generality, let's suppose that the system is prepared in a
state which coincides with an eigenvector of $\hat{A}$, that is to
say, $|\xi\rangle=|a_{j}\rangle$ with $\hat{A}|a_{j}\rangle=a_{j}|a_{j}\rangle$
and $a_{j}\in\mathbb{R}$. In this case it is not difficult verifying
that $\mathrm{Var}(\hat{A})=\mathrm{CovQ}(\hat{A},\hat{B})=\langle[\hat{A},\hat{B}]\rangle=0$.
Therefore the HRSUR gives 
\begin{equation}
0\mathrm{Var}(\hat{B})\ge0.
\end{equation}
So, in this case, the HRSUR does not provide any information about
the possible incompatibility between the observables $\hat{A}$ and
$\hat{B}$. In the next section we shall present the proof of an uncertainty
relation which avoids the triviality problem, witnessing the incompatibility
of two observables even when the system is prepared in one of their
eigenvectors.

\section{Maccone-Pati uncertainty relation}

\label{sec:RIMP}

In contrast to the HRSUR, the Maccone-Pati uncertainty relation (MPUR),
which shall be proved in this section, gives lower bounds for the
sum of the variances associated with two observables \cite{Maccone-Pati}:
\begin{equation}
\mathrm{Var}(\hat{A)}+\mathrm{Var}(\hat{B)}\ge\max(L_{1},L_{2}),\label{eq:RIMP}
\end{equation}
with 
\begin{align}
L_{1} & =2^{-1}|\langle\xi|(\hat{A}\pm\hat{B})|\xi_{\perp}\rangle|^{2},\label{eq:L1}\\
L_{2} & =\pm i\langle[\hat{A},\hat{B}]\rangle+|\langle\xi|(\hat{A}\pm i\hat{B})|\xi_{\perp}\rangle|^{2},\label{eq:L2}
\end{align}
where $|\xi_{\perp}\rangle$ is any normalized vector orthogonal to
the system state $|\xi\rangle$. The signs in Eqs. (\ref{eq:L1})
and (\ref{eq:L2}) are chosen, respectively, to maximize $L_{1}$
and $L_{2}$. Of course, once MPUR holds for any $|\xi_{\perp}\rangle$,
we should search for the $|\xi_{\perp}\rangle$ yielding the bigger
lower bound for the sum of the variances. It is important to note
the the lower bounds $L_{1}$ and $L_{2}$ will be equal to zero only
if the system state, $|\xi\rangle$, is a common eigenvector for both
observables $\hat{A}$ and $\hat{B}$. It is worthwhile mentioning
that the MPUR was already verified experimentally for the special
case of observables represented by unitary operators \cite{Xue}.

It is worthwhile also mentioning that the novelty of MPUR is not simply
the use of the sum of variances instead of their product. One can
easily obtain a HRSUR involving sum of variances by using $(\sigma_{A}-\sigma_{B})^{2}\ge0$,
with the standard deviation of the observable $\hat{X}$ defined as
$\sigma_{X}=\sqrt{\mathrm{Var}(\hat{X})}$. This inequality leads
to 
\begin{equation}
\mathrm{Var}(\hat{A})+\mathrm{Var}(\hat{B})\ge2\sigma_{A}\sigma_{B}\ge|\langle[\hat{A},\hat{B}]\rangle|=T_{2},\label{eq:RI+}
\end{equation}
where the last inequality is a particular case of the HRSUR, Eq. (\ref{eq:RIHRS}).
But one can verify that if the system state is an eigenvector of one
of the observables, then the uncertainty relation of Eq. (\ref{eq:RI+})
also suffers from the triviality problem.

\subsection{Proof of the first lower bound in the MPUR}

\begin{figure}[t]
\begin{centering}
\includegraphics[scale=0.23]{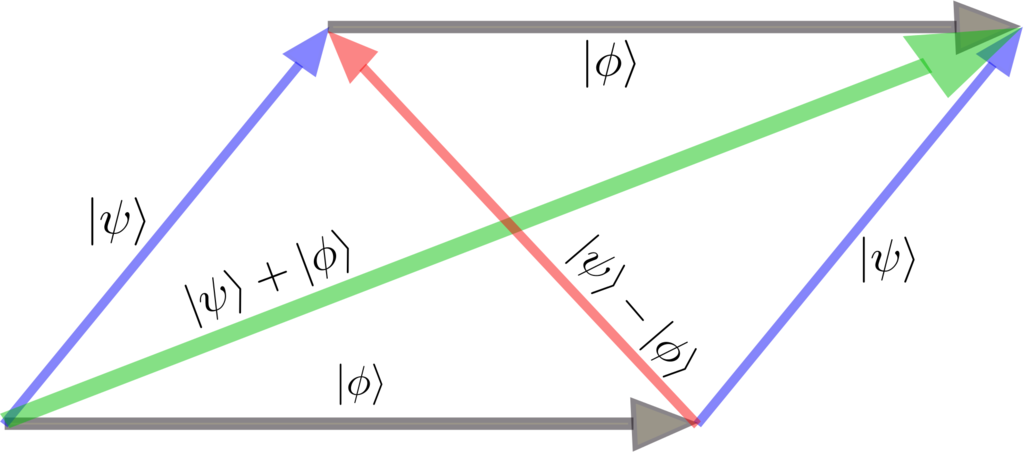}
\par\end{centering}
\caption{Parallelogram law: The sum of the squares of the diagonals of a parallelogram
is equal to the sum of the squares of its sides. In terms of the norms
of the corresponding vectors, this law translates into Eq. (\ref{eq:lei_paralelogramos}).}
\label{fig_lei_paralel}
\end{figure}

For the sake of proving the MPUR, we will make use of parallelogram
law. This rule is depicted in Fig. \ref{fig_lei_paralel} and states
that for any two vectors $|\psi\rangle$ and $|\phi\rangle$ in the
Hilbert space $\mathcal{H}$, the following equality holds: 
\begin{equation}
2({\normalcolor \parallel}|\psi\rangle{\normalcolor \parallel}^{2}+{\normalcolor \parallel}|\phi\rangle{\normalcolor \parallel}^{2})={\normalcolor \parallel}(|\psi\rangle+|\phi\rangle){\normalcolor \parallel}^{2}+{\normalcolor \parallel}(|\psi\rangle-|\phi\rangle){\normalcolor \parallel}^{2}.\label{eq:lei_paralelogramos}
\end{equation}

Let us insert the vectors defined in Eq. (\ref{eq:states}) in the
parallelogram law, Eq. (\ref{eq:lei_paralelogramos}). As ${\normalcolor \parallel}|\psi\rangle{\normalcolor \parallel}^{2}=\mathrm{Var}(\hat{A})$
and ${\normalcolor \parallel}|\phi\rangle{\normalcolor \parallel}^{2}=\mathrm{Var}(\hat{B})$
we shall have 
\begin{eqnarray}
 &  & \mathrm{Var}(\hat{A})+\mathrm{Var}(\hat{B})\nonumber \\
 &  & =2^{-1}({\normalcolor \parallel}(|\psi\rangle+|\phi\rangle){\normalcolor \parallel}^{2}+{\normalcolor \parallel}(|\psi\rangle-|\phi\rangle){\normalcolor \parallel}^{2})\label{eq:a}\\
 &  & \ge2^{-1}{\normalcolor \parallel}(|\psi\rangle\pm|\phi\rangle){\normalcolor \parallel}^{2}\label{eq:b}\\
 &  & =2^{-1}(\langle\psi|\pm\langle\phi|)(|\psi\rangle\pm|\phi\rangle)\langle\xi_{\perp}|\xi_{\perp}\rangle\label{eq:c}\\
 &  & \ge2^{-1}|(\langle\psi|\pm\langle\phi|)|\xi_{\perp}\rangle|^{2}\label{eq:d}\\
 &  & =2^{-1}|\langle\xi|(\hat{A}\pm\hat{B})|\xi_{\perp}\rangle-(\langle\hat{A}\rangle\pm\langle\hat{B}\rangle)\langle\xi|\xi_{\perp}\rangle|^{2}\label{eq:e}\\
 &  & =2^{-1}|\langle\xi|(\hat{A}\pm\hat{B})|\xi_{\perp}\rangle|^{2}=L_{1}.\label{eq:f}
\end{eqnarray}
We obtained the inequality in Eq. (\ref{eq:b}) from the equality
in Eq. (\ref{eq:a}) by applying the positivity of the norm. We get
from (\ref{eq:b}) to (\ref{eq:c}) and from (\ref{eq:e}) to (\ref{eq:f})
using a normalized vector $|\xi_{\perp}\rangle$ which is orthogonal
to the system state $|\xi\rangle$. By its turn, the inequality of
Eq. (\ref{eq:d}) is a consequence of the Cauchy-Schwarz inequality,
Eq. (\ref{eq:DCS}). The signs in the equations above depend on if
${\normalcolor \parallel}(|\psi\rangle+|\phi\rangle){\normalcolor \parallel}^{2}$
or ${\normalcolor \parallel}(|\psi\rangle-|\phi\rangle){\normalcolor \parallel}^{2}$
is used when going from Eq. (\ref{eq:a}) to Eq. (\ref{eq:b}) and
are chosen to maximize $L_{1}$.

\subsection{Proof of the second lower bound in the MPUR}

By applying the same procedures of the last sub-section, we can verify
that 
\begin{eqnarray}
{\normalcolor \parallel}(|\psi\rangle\pm i|\phi\rangle){\normalcolor \parallel}^{2} & = & (\langle\psi|\mp i\langle\phi|)(|\psi\rangle\pm i|\phi\rangle)\nonumber \\
 & = & {\normalcolor \parallel}|\psi\rangle{\normalcolor \parallel}^{2}+{\normalcolor \parallel}|\phi\rangle{\normalcolor \parallel}^{2}\pm i(\langle\psi|\phi\rangle-\langle\phi|\psi\rangle)\nonumber \\
 & = & {\normalcolor \parallel}|\psi\rangle{\normalcolor \parallel}^{2}+{\normalcolor \parallel}|\phi\rangle{\normalcolor \parallel}^{2}\pm i\langle[\hat{A},\hat{B}]\rangle
\end{eqnarray}
and 
\begin{eqnarray}
{\normalcolor \parallel}(|\psi\rangle\pm i|\phi\rangle){\normalcolor \parallel}^{2} & = & (\langle\psi|\mp i\langle\phi|)(|\psi\rangle\pm i|\phi\rangle)\langle\xi_{\perp}|\xi_{\perp}\rangle\nonumber \\
 & \ge & |(\langle\psi|\mp i\langle\phi|)|\xi_{\perp}\rangle|^{2}\nonumber \\
 & = & |\langle\xi|(\hat{A}\mp i\hat{B})|\xi_{\perp}\rangle-(\langle\hat{A}\rangle\mp i\langle\hat{B}\rangle)\langle\xi|\xi_{\perp}\rangle|^{2}\nonumber \\
 & = & |\langle\xi|(\hat{A}\mp i\hat{B})|\xi_{\perp}\rangle|^{2}.\label{eq:ineq}
\end{eqnarray}

Thus, if we utilize $i|\phi\rangle$ in place of $|\phi\rangle$ in
the parallelogram law, as ${\normalcolor \parallel}i|\phi\rangle{\normalcolor \parallel}={\normalcolor \parallel}|\phi\rangle{\normalcolor \parallel}$,
we get 
\begin{eqnarray}
2(\mathrm{Var}(\hat{A})+\mathrm{Var}(\hat{B})) & \ge & \mathrm{Var}(\hat{A})+\mathrm{Var}(\hat{B})\pm i\langle[\hat{A},\hat{B}]\rangle\nonumber \\
 &  & +|\langle\xi|(\hat{A}\pm i\hat{B})|\xi_{\perp}\rangle|^{2},\label{eq:ineq2}
\end{eqnarray}
from which we promptly obtain the lower bound $L_{2}$ of Eq. (\ref{eq:L2}).
The sign in Eq. (\ref{eq:ineq2}) is determined by which of the terms
${\normalcolor \parallel}(|\psi\rangle\pm i|\phi\rangle){\normalcolor \parallel}^{2}$
in Eq. (\ref{eq:lei_paralelogramos}) the inequality (\ref{eq:ineq})
is applied to, and is chosen such that $L_{2}$ is maximized.

\section{Example: Complementarity for a qubit}
\label{sec:exemplos}

\begin{figure}[t]
\begin{centering}
\includegraphics[scale=0.23]{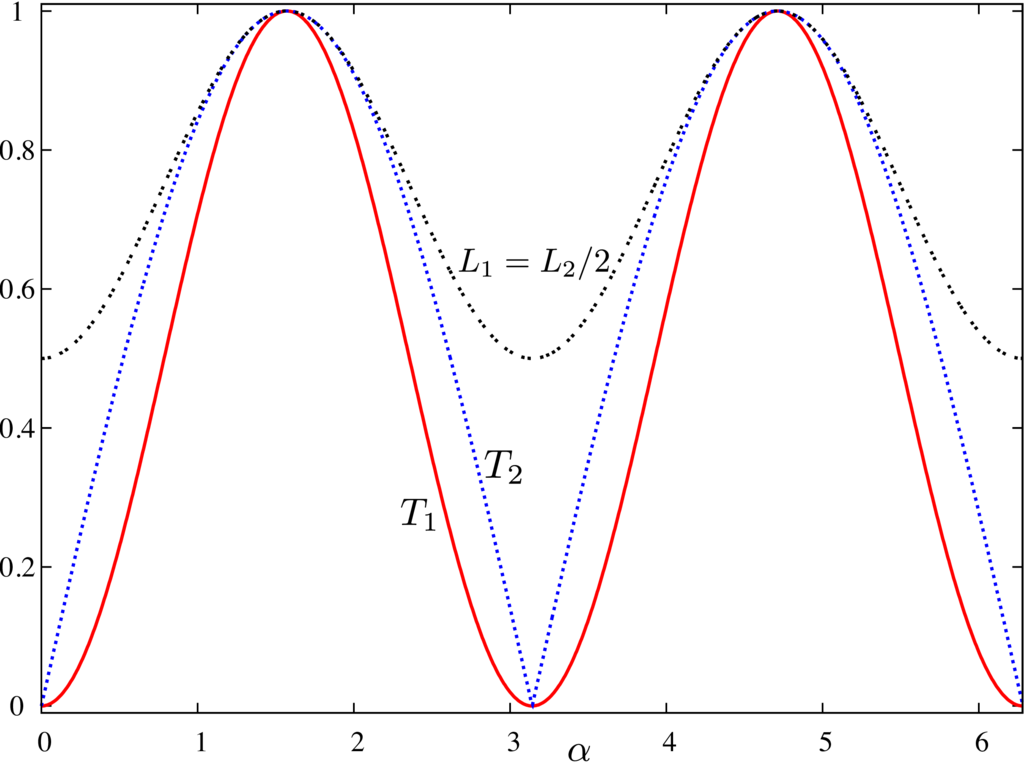}
\par\end{centering}
\caption{Lower bounds for the variances for the observables $\hat{X}$ and
$\hat{Z}$, as imposed by the uncertainty relations of Heisenberg,
Robertson, and Schrödinger ($T_{1}$ e $T_{2}$) and of Maccone and
Pati ($L_{1}$ e $L_{2}$), for a qubit prepared in state $|\xi\rangle=2^{-1/2}(|0\rangle+e^{i\alpha}|1\rangle)$.}
\label{fig:qubit}
\end{figure}

In this section we look at a two-level system, a qubit, prepared in
the state 
\begin{equation}
|\xi\rangle=2^{-1/2}(|0\rangle+e^{i\alpha}|1\rangle),
\end{equation}
with $|0\rangle$ and $|1\rangle$ being eigenvectors of the Pauli
matrix $\hat{Z}=|0\rangle\langle0|-|1\rangle\langle1|$ and $\alpha\in[0,2\pi)$.
Of course, everything we say in this section holds for the popular
example of a spin 1/2 particle measured with Stern-Gerlach apparatuses
\cite{Sakurai}. We regard the application of HRSUR and MPUR to witness
the well known incompatibility between the observables $\hat{Z}$
and $\hat{X}=|0\rangle\langle1|+|1\rangle\langle0|$. One can verify
that for the state $|\xi\rangle$: $\langle\hat{Z}\rangle=0$, $\langle\hat{X}\rangle=\cos\alpha$,
and $\langle\hat{Z}\hat{X}\rangle=-\langle\hat{X}\hat{Z}\rangle=i\sin\alpha$.
We this we have $\mathrm{CovQ}(\hat{X},\hat{Z})=0$ e $|\langle[\hat{X},\hat{Z}]\rangle|^{2}=2^{2}\sin^{2}\alpha$.
The two lower bounds in the HRSUR of Eqs. (\ref{eq:RIHRS}) and (\ref{eq:RI+})
are then given by 
\begin{equation}
T_{1}=2^{-2}T_{2}^{2}=\sin^{2}\alpha.
\end{equation}
Taking into account that for this example there is only one normalized
vector orthogonal to $|\xi\rangle$: $|\xi_{\perp}\rangle=2^{-1/2}(|0\rangle-e^{i\alpha}|1\rangle)$,
after some simple calculations, we obtain the lower bounds for the
MPUR, Eqs. (\ref{eq:L1}) e (\ref{eq:L2}): 
\begin{equation}
L_{2}=2L_{1}=1+\sin^{2}\alpha.
\end{equation}
These four lower bounds for the variances of $\hat{X}$ and $\hat{Z}$
are shown in Fig. \ref{fig:qubit}. We see that even though the qualitative
behavior of the curves is generally similar, there are important quantitative
differences for the phases $\alpha=\{0,\pi,2\pi\}$. For these values
of $\alpha$, the system state, $|\xi\rangle$, is an eigenvector
of $\hat{X}$ and, in contrast to the MPUR, the HRSUR, due to the
triviality problem, is not capable of indicating that the width of
the probability distribution for the eigenvalues of $\hat{Z}$ is
non-null.

\section{Final remarks}

\label{sec:conclusoes}

In this article, after discussing some aspects of the uncertainty
principle of quantum mechanics (QM), we presented a didactic proof
of the Maccone-Pati uncertainty relation and exemplified its application
to a two-level system. It is a curious fact that a relevant restriction
within QM (as is the Heisenberg-Robertson-Schödinger uncertainty relation)
has an important problem which, although probably being for long noticed
by several teachers and researchers in the area, was solved so much
time after its conception. Thus we hope that the simple derivation
of the MPUR we presented in this article will further motivate its
inclusion in QM courses.
\begin{acknowledgments}
This work was supported by CNPq, processes 441875/2014-9 and 303496/2014-2,
by the Instituto Nacional de Ciência e Tecnologia de Informação Quântica
(INCT-IQ), process 2008/57856-6, and by CAPES, process 6531/2014-08. \end{acknowledgments}

\end{document}